\documentclass[3p,times,procedia]{elsarticle}
\usepackage{nupha_ecrc}

\volume{00}

\firstpage{1}

\journalname{Nuclear Physics A}

\runauth{A. Ramamurti, E. Shuryak}

\jid{nupha}

\jnltitlelogo{Nuclear Physics A}

\usepackage{amsmath}
\usepackage{tipa}
\usepackage{amssymb}
\usepackage{graphicx}
\usepackage{xfrac}
\usepackage{bm}
\usepackage{enumerate}
\usepackage{color}
\usepackage[colorlinks = true,
            linkcolor = blue,
            urlcolor  = blue,
            citecolor = blue,
            anchorcolor = blue]{hyperref}
\usepackage{subfigure}
\usepackage{braket}
\usepackage{capt-of}
\usepackage{chngcntr}

\newcommand{\eq}[1]{\begin{equation}#1\end{equation}}

\newcommand{\Tr}{\mbox{Tr}\;}

\begin{document}

\begin{frontmatter}



\dochead{XXVIth International Conference on Ultrarelativistic Nucleus-Nucleus Collisions\\ (Quark Matter 2017)}

\title{An Effective Model of QCD Monopoles}

\author{Adith Ramamurti \fnref{a}}
\author{Edward Shuryak}

\address{Department of Physics and Astronomy, Stony Brook University, Stony Brook, New York 11794-3800, USA}

\fntext[a]{Speaker: adith.ramamurti@stonybrook.edu}

\begin{abstract}

In this work, we carried out quantum many-body studies of magnetic monopole ensembles through numerical simulations of the path integral for one- and two-component Coulomb Bose systems.  We found the relation between the critical temperature for the Bose-Einstein condensation phase transition and the Coulomb coupling strength  using two methods, the finite-size scaling of the superfluid fraction and statistical analysis of permutation cycles.  After finding parameters that match the correlation functions measured in our system with the correlation functions previously measured on the lattice, we arrived at an effective quantum model of color magnetic monopoles in QCD.  From this matched model, we were able to extract the monopole contribution to QCD equation of state near $T_\text{c}$. This proceeding and the given talk are based on \cite{Ramamurti:2017fdn}.

\end{abstract}
\end{frontmatter}

\section{Introduction}

In the 1970s, Nambu \cite{Nambu:1974zg}, `t Hooft \cite{NUPHA.B190.455}, and Mandelstam \cite{Mandelstam:1974pi} proposed that that confinement is due to the Bose-Einstein condensation (BEC) of magnetic monopoles. Based on this model, the ``magnetic scenario" for finite-temperature QCD \cite{Liao:2006ry} suggested that interactions of the ``electric" objects -- quarks and gluons -- with magnetic monopoles give rise to the unusual transport properties of the quark-gluon plasma (QGP).

Furthermore, lattice simulations of gauge theories found magnetic monopoles, and were able to measure their spatial correlation functions and densities \cite{DAlessandro:2007lae,D'Alessandro:2010xg,Bonati:2013bga}. These simulations, at $T>T_\text{c}$, also include many more degrees of freedom, such as quark and gluon quasiparticles, which make it difficult to isolate the thermodynamic contributions of the monopoles. In creating an effective quantum model, we aimed to match the results found in these lattice simulations and identify the contributions of the monopoles to observables.

\section{Numerical Methods}

To simulate quantum systems of particles, we used Path-Integral Monte Carlo (PIMC). Following Feynman \cite{Feynman:1948ur,PhysRev.91.1291,Feynman:1953zz}, each particle is represented by a periodic path in imaginary time. From these paths, we can construct the density matrix $\rho$ for the system and, thus, the partition function, which is given by $Z = \Tr[\rho]$. In the PIMC method, the particles are put in a periodic box and the configurations of the imaginary-time paths are sampled numerically using Metropolis Monte Carlo, with the configuration weight given by the Euclidean action of the paths. From these configurations, we can construct the density matrix and partition function, and therefore the thermodynamics, of the system. For an overview of the method, see \cite{RevModPhys.67.279}.

In addition to studying the partition function of the system, the path configurations can be used to study the BEC transition. In the path integral formalism, the paths of bosonic particles can be {\em permuted}. At the critical temperature, the suppression of these permutations disappears, and the imaginary-time paths wrap around the periodic box in both the spatial and temporal directions. Using two methods, we can use the statistics of these windings to identify the critical temperature for condensation.

The first method utilizes the finite size scaling of the superfluid fraction of the system \cite{PhysRevB.46.3535}. The superfluid fraction can be related to the winding number by
\eq{
\frac{\rho_s}{\rho}(T) \propto \frac{\braket{W^2(T)}}{2 \beta N} \,,
}
where $W$ is the spatial winding number, $\beta$ is the inverse temperature, and $N$ is the number of particles in the system. At infinite volume, one expects that below $T_\text{c}$ the superfluid fraction will be 1, due to the disappearance of suppression of permutations, and 0 above $T_\text{c}$. At finite volume, this jump is not as sharp, as one does not need an infinite permutation to wrap around the box. By studying the finite-size scaling of this transition with the system size, one can identify the critical temperature.

In addition to spatial winding, the temporal windings can be used to identify $T_\text{c}$ \cite{D'Alessandro:2010xg}. From study of the partition function of a Bose gas, we expect that the density of $k$-cycles is given by
\eq{
\rho_k(T) = \exp(-\hat{\mu}(T)\cdot k)/k^\gamma \,,
}
where $k$ is the number of times a path winds around in the temporal direction and $\gamma$ a non-negative real number. $T_\text{c}$ is found from the condition that the exponential suppression $\hat{\mu}(T) \rightarrow 0$ at $T \rightarrow T_\text{c}$.

\section{Critical Temperature of One- and Two-Component Coulomb Bose Gases}

Classical simulations \cite{Liao:2006ry,PhysRevLett.101.162302} showed that the magnetic monopoles acted as a Coulombic liquid with a magnetic coupling that increases with temperature. To study the effect of different couplings in quantum Coulomb Bose systems, we simulated systems of one- and two-component Bose particles interacting via potential $V(r) \propto \alpha/r$, and varied the coupling $\alpha$.

\begin{figure}[h!]
\begin{center}
\subfigure[]{
\includegraphics[width=.37\columnwidth,angle=0]{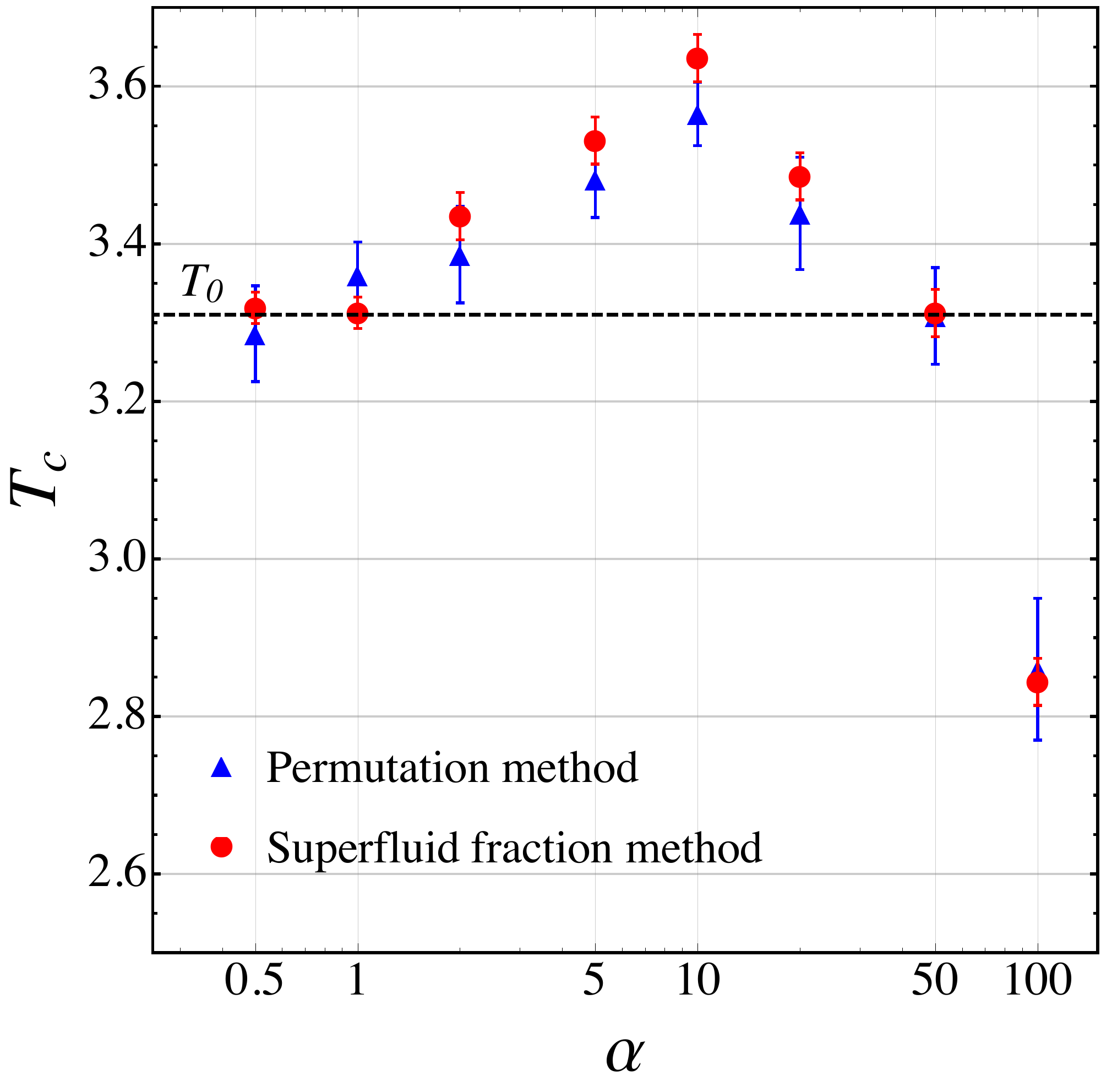}
}
\subfigure[]{
\includegraphics[width=.37\columnwidth,angle=0]{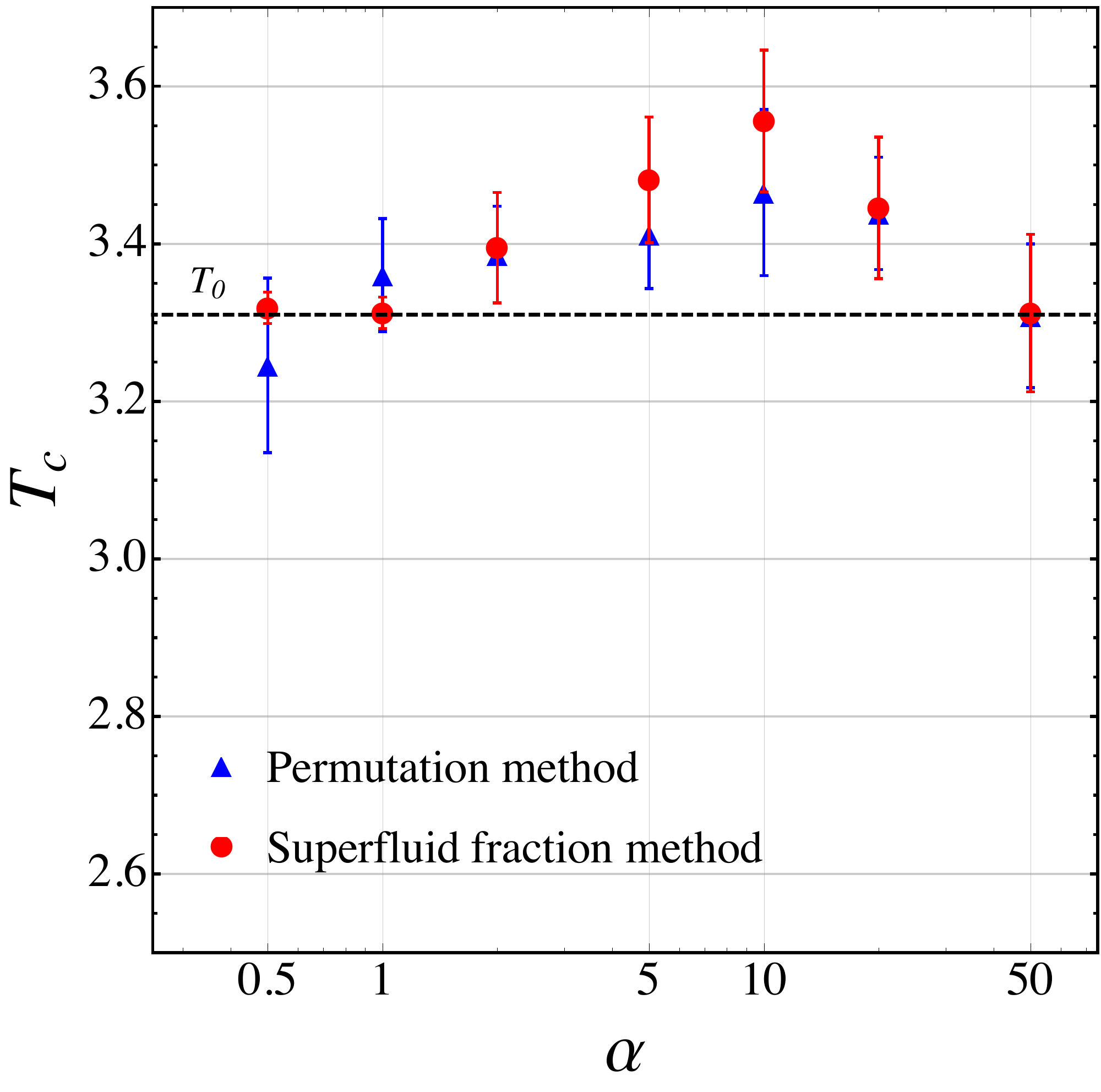}
}
\caption{The critical temperature for the BEC phase transition as a function of the coupling, $\alpha$, for (a) one-component and (b) two-component Coulomb Bose systems. The red circles are the results of the finite-size scaling superfluid fraction calculation, and the blue triangles are the results of the permutation-cycle calculation for a system with 32 particles. The black dashed line denotes the Einstein ideal Bose gas critical temperature, $T_0$.}
\label{fig:crit_temp}
\end{center}
\end{figure}

The results of our simulations for one- and two-component Coulomb Bose gases are shown in Fig. \ref{fig:crit_temp}. We see that for both cases, as the coupling is increased $T_\text{c}$ rises above the ideal gas critical temperature, $T_0$, before falling back to and below $T_0$.

\section{Parameters for an Effective Model of Magnetic Monopoles}
\begin{figure}[h!]
\begin{center}
\subfigure[]{
\includegraphics[width=.37\columnwidth,angle=0]{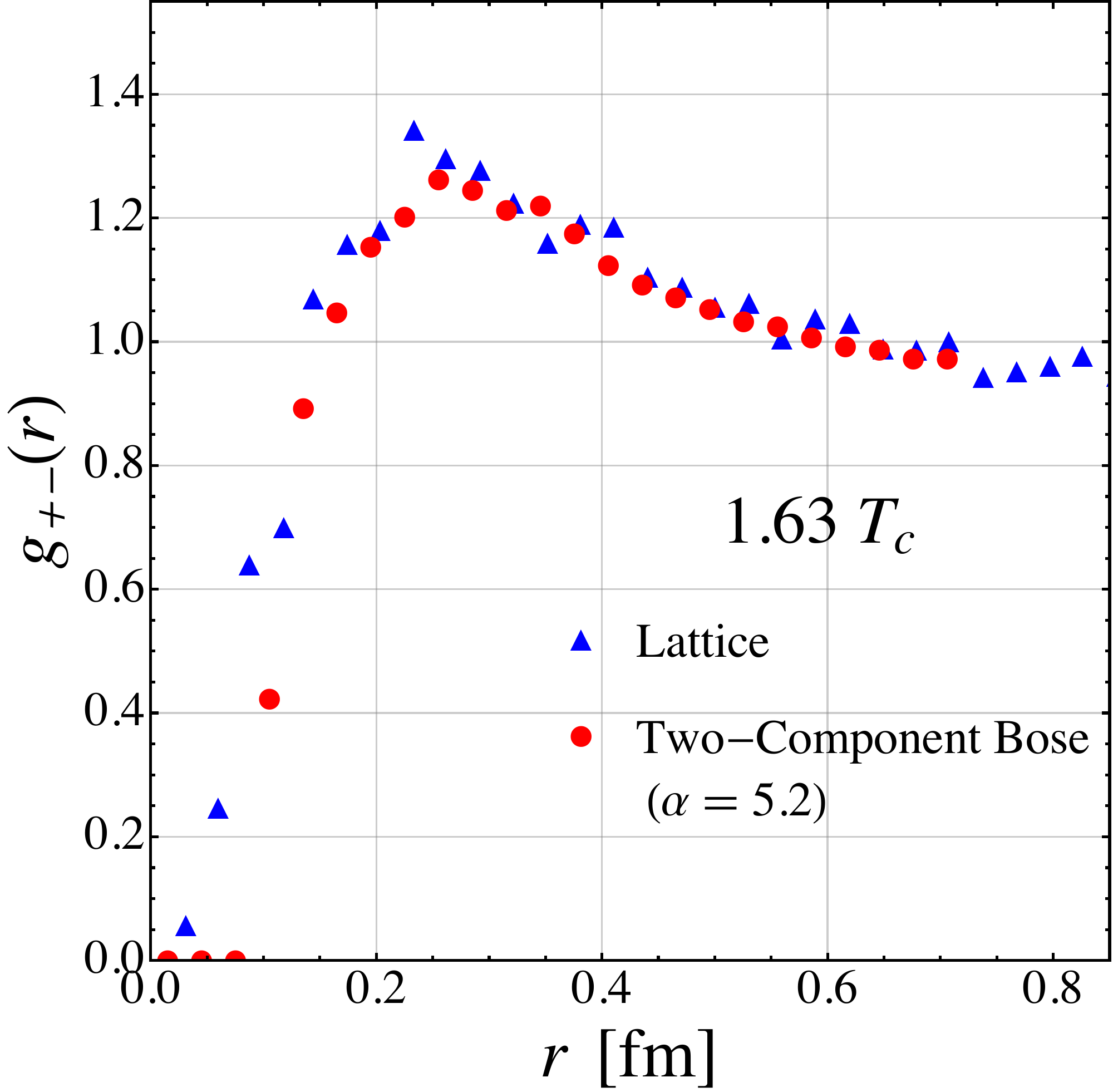}
}
\subfigure[]{
\includegraphics[width=.37\columnwidth,angle=0]{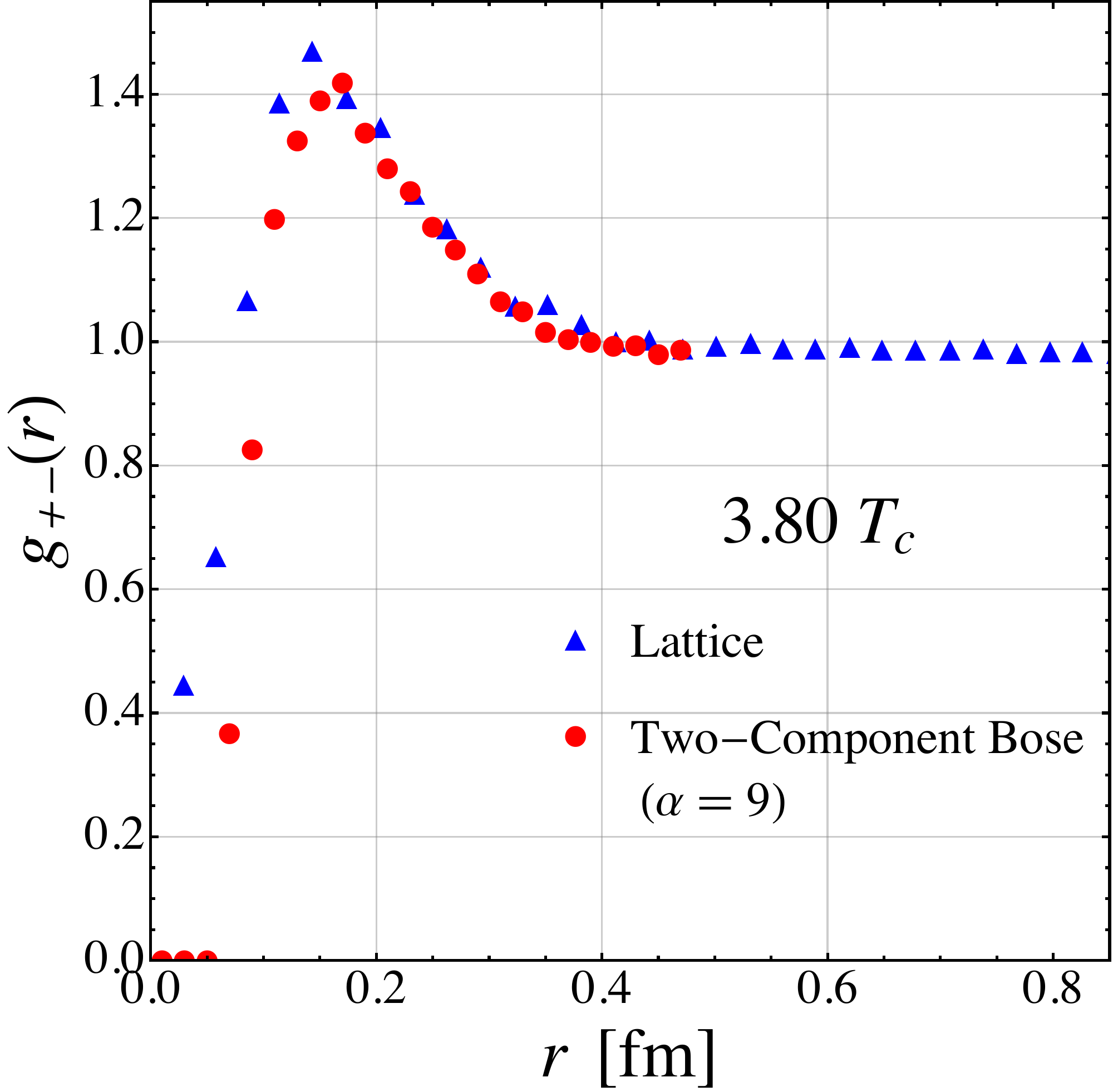}
}
\caption{Matched radial distribution functions for (a) 1.63$T_\text{c}$ and (b) 3.80 $T_\text{c}$. The red circles are from PIMC simulations, and the blue triangles are from SU(2) lattice simulations \cite{DAlessandro:2007lae}. }
\label{fig:lat_match}
\end{center}
\end{figure}

From our path configurations, we are able to measure the spatial correlations of monopoles and antimonopoles. To find the parameters that correspond to physical monopoles, we compared and matched our spatial correlations to those found in pure-gauge SU(2) lattice simulations \cite{DAlessandro:2007lae}. Two examples of these matched correlation functions are shown in Fig. \ref{fig:lat_match}. We found that, to reproduce lattice results, the temperature-dependent coupling for a two-component Coulomb Bose gas with unit density is given by 
\eq{\alpha(T/T_\text{c}) \approx 3.4\, \rho_m^{1/3}(T/T_\text{c})\,,} 
where $\alpha(T)$ the coupling used in our simulation, $\rho_m(T)$ the monopole density (in fm$^{-3}$) found in \cite{DAlessandro:2007lae}.

\section{Contribution of Magnetic Monopoles to Thermodynamics}

\begin{figure}[h!]
\begin{center}
\includegraphics[width=.47\textwidth,angle=0]{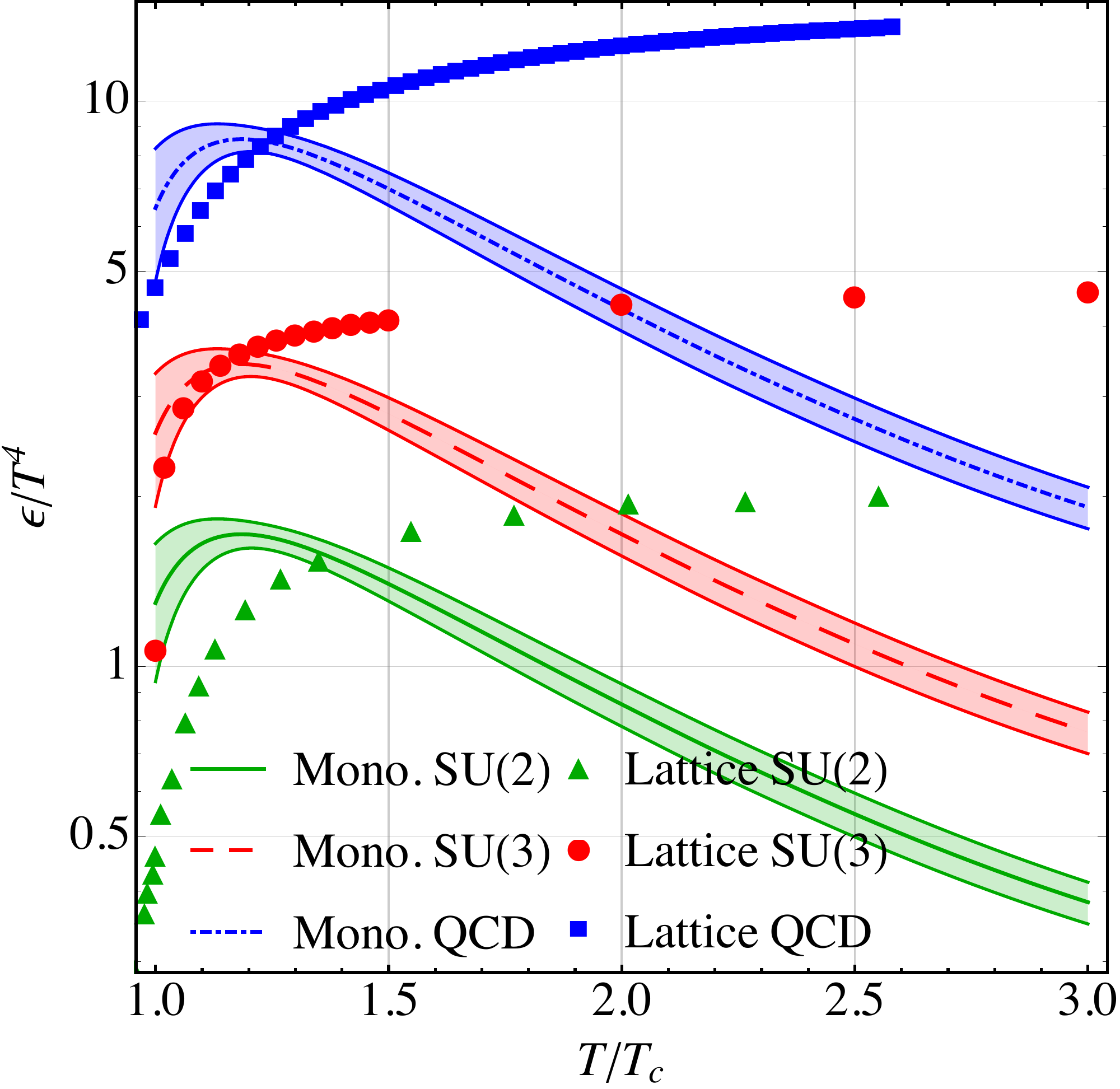}
\caption{The energy density of the monopoles compared to lattice data for pure-gauge SU(2) and SU(3). Lattice SU(2) results are from \cite{Engels:1994xj,Engels:1990vr}, SU(3) from \cite{Borsanyi:2012ve}, and QCD from \cite{Bazavov:2014pvz}.}
\label{fig:latencomp}
\end{center}
\end{figure}

Taking into account the mass of the particles, which has been extracted from the lattice in \cite{D'Alessandro:2010xg}, we show in Fig. \ref{fig:latencomp} the energy density contributions of the monopoles compared to the total energy density found on the lattice.  One can see that the monopoles contribute a significant portion of the total energy just above $T_\text{c}$. 

The lattice data for the SU(2) color group can be directly compared to our study. For the SU(3) color group, on the other hand, we have estimated the monopole contribution to the energy density by doubling the energy density found in our study, since SU(3) has two diagonal color generators and therefore has two distinct species of monopoles. This scaling does not take into account the energy coming from the interaction between the two species, which will have to be studied further in the future. Finally, to go from SU(3) to QCD, from standard counting of states, one finds that the number of quark-monopole states for two light flavors also increases the number of species by the factor of 3. 

\section{Summary}
In this work, we have studied the effects of Coulomb interaction on the Bose-Einstein condensation. We numerically calculated the critical temperature of one- and two-component Coulomb systems, by two different methods, as a function of the interaction strength. We then mapped the results of the two-component case to the results of lattice simulations of color magnetic monopoles in pure-gauge SU(2), and found a very good agreement. Finally, we determined the  monopole contribution  to the overall thermodynamics (energy density) of the thermal matter, at and above $T_\text{c}$ in pure-gauge SU(2), and made estimates for SU(3) and QCD theories. We conclude that the monopoles possibly dominate the thermodynamics of these theories just above $T_\text{c}$.

This work was supported in part by the U.S. D.O.E. Office of Science, Contract No. DE-FG-88ER40388.
 
\bibliographystyle{elsarticle-num}

\begin{thebibliography}{00}
 
\bibitem{Ramamurti:2017fdn} 
  A.~Ramamurti and E.~Shuryak,
Phys.\ Rev.\ D, in press (2017).
  arXiv:1702.07723 [hep-ph].

 
\bibitem{Nambu:1974zg} 
  Y.~Nambu,
  Phys.\ Rev.\ D {\bf 10}, 4262 (1974).
  doi:10.1103/PhysRevD.10.4262

\bibitem{NUPHA.B190.455} 
  G.~`t Hooft,
  Nucl.\ Phys.\ B {\bf 190}, 455 (1981).
  doi:10.1016/0550-3213(81)90442-9

\bibitem{Mandelstam:1974pi} 
  S.~Mandelstam,
  Phys.\ Rept.\  {\bf 23}, 245 (1976).
  doi:10.1016/0370-1573(76)90043-0


\bibitem{Liao:2006ry} 
  J.~Liao and E.~Shuryak,
  Phys.\ Rev.\ C {\bf 75}, 054907 (2007)
  doi:10.1103/PhysRevC.75.054907
  
\bibitem{DAlessandro:2007lae} 
  A.~D'Alessandro and M.~D'Elia,
  Nucl.\ Phys.\ B {\bf 799}, 241 (2008)
  doi:10.1016/j.nuclphysb.2008.03.002
    
\bibitem{D'Alessandro:2010xg} 
  A.~D'Alessandro, M.~D'Elia, and E.~V.~Shuryak,
  Phys.\ Rev.\ D {\bf 81}, 094501 (2010)
  doi:10.1103/PhysRevD.81.094501

\bibitem{Bonati:2013bga} 
  C.~Bonati and M.~D'Elia,
  Nucl.\ Phys.\ B {\bf 877}, 233 (2013)
  doi:10.1016/j.nuclphysb.2013.10.004


\bibitem{Feynman:1948ur} 
  R.~P.~Feynman,
  Rev.\ Mod.\ Phys.\  {\bf 20}, 367 (1948).
  doi:10.1103/RevModPhys.20.367

\bibitem{PhysRev.91.1291} 
  R.~P.~Feynman,
  Phys.\ Rev.\  {\bf 91}, 1291 (1953).
  doi:10.1103/PhysRev.91.1291

\bibitem{Feynman:1953zz} 
  R.~P.~Feynman,
  Phys.\ Rev.\  {\bf 91}, 1301 (1953).
  doi:10.1103/PhysRev.91.1301


\bibitem{RevModPhys.67.279} 
  D.~M.~Ceperley,
  Rev.\ Mod.\ Phys.\  {\bf 67}, 279 (1995).
  doi:10.1103/RevModPhys.67.279
  
  \bibitem{PhysRevB.46.3535}
	E.~L.~Pollock and K.~J.~Runge,
	Phys.\ Rev.\ B {\bf 46}, 3535 (1992)
	doi:10.1103/PhysRevB.46.3535


\bibitem{PhysRevLett.101.162302} 
  J.~Liao and E.~Shuryak,
  Phys.\ Rev.\ Lett.\  {\bf 101}, 162302 (2008)
  doi:10.1103/PhysRevLett.101.162302

\bibitem{Engels:1994xj} 
  J.~Engels, F.~Karsch and K.~Redlich,
  Nucl.\ Phys.\ B {\bf 435}, 295 (1995)
  doi:10.1016/0550-3213(94)00491-V
  
\bibitem{Engels:1990vr} 
  J.~Engels, J.~Fingberg, F.~Karsch, D.~Miller and M.~Weber,
  Phys.\ Lett.\ B {\bf 252}, 625 (1990).
  doi:10.1016/0370-2693(90)90496-S
  
\bibitem{Borsanyi:2012ve} 
  S.~Borsanyi, G.~Endrodi, Z.~Fodor, S.~D.~Katz and K.~K.~Szabo,
  JHEP {\bf 1207}, 056 (2012)
  doi:10.1007/JHEP07(2012)056
 
\bibitem{Bazavov:2014pvz} 
  A.~Bazavov {\it et al.} [HotQCD Collaboration],
  Phys.\ Rev.\ D {\bf 90}, 094503 (2014)
  doi:10.1103/PhysRevD.90.094503

 \end{thebibliography}


\end{document}